%% file: ANA-EXOT-2017-17-PAPER.tex
\pdfoutput=1
\newcommand*{\ATLASLATEXPATH}{}
\documentclass[cernpreprint, texlive=2017, USenglish, english, block=none]{\ATLASLATEXPATH atlasdoc}
\usepackage[subfigure]{\ATLASLATEXPATH atlaspackage}
\usepackage{\ATLASLATEXPATH atlasbiblatex}
 
\usepackage{\ATLASLATEXPATH atlasphysics}
 
\graphicspath{{logos/}{figures/}}
 
\usepackage{ANA-EXOT-2017-17-PAPER-defs}

\usepackage{float}
\usepackage{multirow}
\usepackage{tikz}
\usepackage{amssymb}
\usepackage{lscape}
\usepackage{array}
\usepackage{tabulary}
\newcolumntype{K}[1]{>{\centering\arraybackslash}p{#1}}
\usepackage{amssymb}
\usepackage{pifont}

\AtlasTitle{Combination of the searches for pair-produced vector-like partners of the third-generation quarks at $\sqrt{s} =$~13~TeV with the ATLAS detector}

\AtlasAbstract{
A combination of the searches for pair-produced vector-like partners
of the top and bottom quarks in various decay channels
($T$$\rightarrow$$Zt/Wb/Ht$, $B$$\rightarrow$$Zb/Wt/Hb$) is performed
using 36.1~fb$^{-1}$ of $pp$ collision data at $\sqrt{s}$ = 13~TeV
with the ATLAS detector at the Large Hadron Collider. The observed
data are found to be in good agreement with the Standard Model
background prediction in all individual searches. Therefore, combined
95\% confidence-level upper limits are set on the production
cross-section for a range of vector-like quark scenarios,
significantly improving upon the reach of the individual searches.
Model-independent limits are set assuming the vector-like quarks decay
to Standard Model particles.  A singlet $T$ is excluded for masses
below 1.31~TeV and a singlet $B$ is excluded for masses below
1.22~TeV.  Assuming a weak isospin $(T,B)$ doublet and $|V_{Tb}| \ll
|V_{tB}|$, $T$ and $B$ masses below 1.37~TeV are excluded.  }

\AtlasRefCode{EXOT-2017-17}
 
\PreprintIdNumber{CERN-EP-2018-205}

\arXivId{1808.02343}
 
\HepDataRecord{https://www.hepdata.net/record/83541}
 
\AtlasJournalRef{\PRL 121 (2018) 211801}
\AtlasDOI{10.1103/PhysRevLett.121.211801}

\hypersetup{pdftitle={ATLAS document},pdfauthor={The ATLAS Collaboration}}

\begin{document}
 
\maketitle
 
\setlength{\columnsep}{7mm}

\section{Introduction}
 
Naturalness arguments~\cite{Susskind:1978ms} suggest there should be a
mechanism that cancels out the quadratically-divergent contributions to
the Higgs boson mass caused by radiative corrections from Standard
Model (SM) particles.  Several explanations are proposed in theories
beyond the SM.
Little Higgs~\cite{ArkaniHamed:2002qy,Schmaltz:2005ky} and Composite
Higgs~\cite{Kaplan:1983sm,Agashe:2004rs} models introduce a
spontaneously-broken global symmetry, with the Higgs boson emerging as
a pseudo Nambu--Goldstone boson~\cite{Hill:2002ap}. Such models
predict the existence of vector-like quarks (VLQs), color-triplet
spin-1/2 fermions whose left- and right-handed chiralities transform in the
same way under
weak-isospin~\cite{delAguila:1982fs,AguilarSaavedra:2009es}. In these
models, VLQs are expected to couple preferentially to third-generation
quarks~\cite{delAguila:1982fs,Aguilar-Saavedra:2013wba} and can have
flavor-changing neutral-current decays in addition to charged-current
decays.  An up-type VLQ $T$ with charge $+2/3$ can decay into $Wb$,
$Zt$, or $Ht$. Similarly, a down-type quark $B$ with charge $-1/3$ can
decay into $Wt$, $Zb$, or $Hb$.  In order to be consistent with results from
precision electroweak measurements, the mass-splitting between
VLQs belonging to the same SU(2) multiplet is
required to be small~\cite{Aguilar-Saavedra:2013qpa}, forbidding cascade decays
such as $T$$\to$$WB$. Couplings between the VLQs and the first- and
second-generation quarks, although not favored, are not
excluded~\cite{Atre:2008iu,Atre:2011ae}.
 
\begin{figure}[h!]
\begin{center}
\centering
\subfigure[]{\includegraphics[width=0.45\textwidth,height=0.24\textwidth,keepaspectratio]{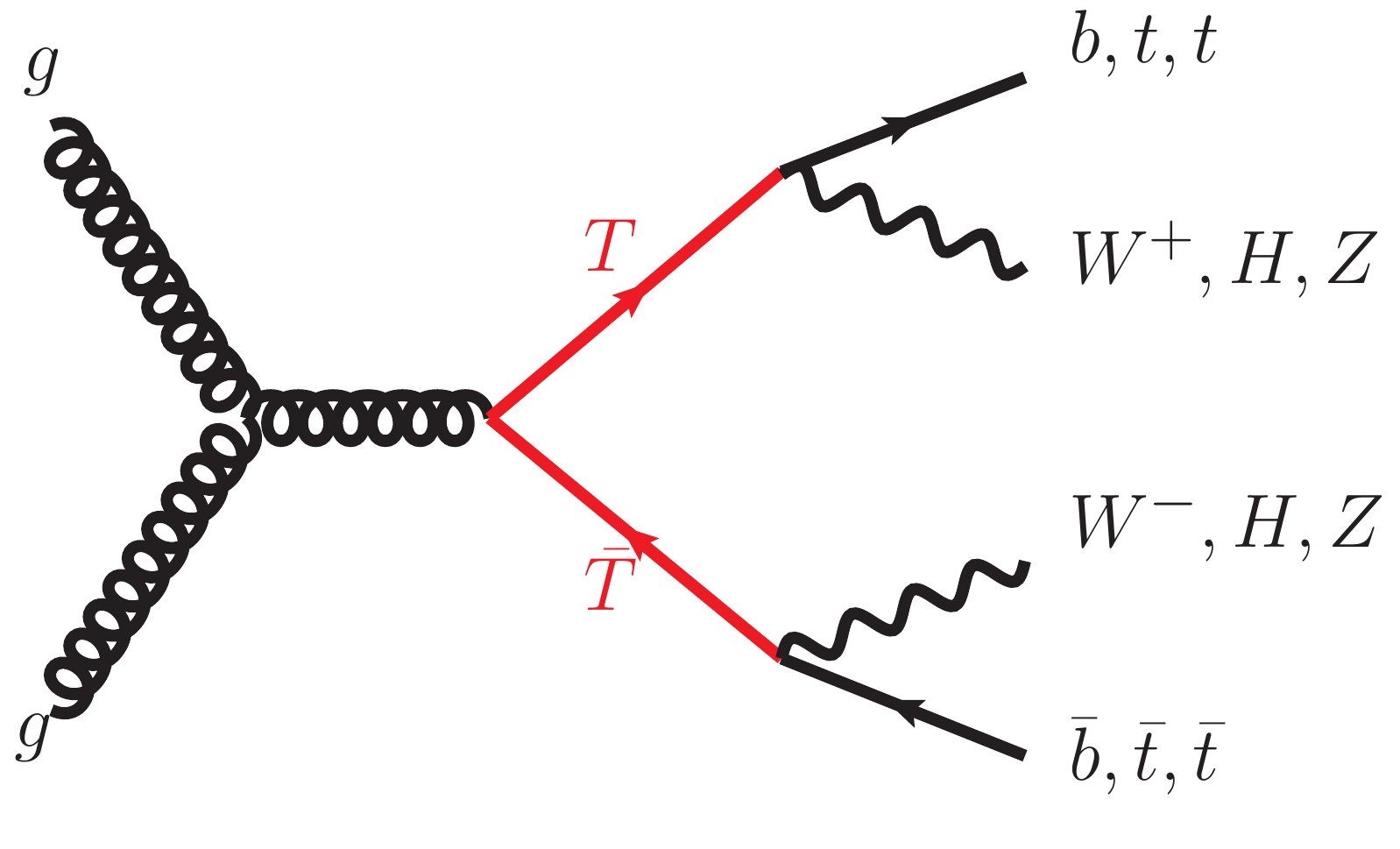}}
\subfigure[]{\includegraphics[width=0.45\textwidth,height=0.24\textwidth,keepaspectratio]{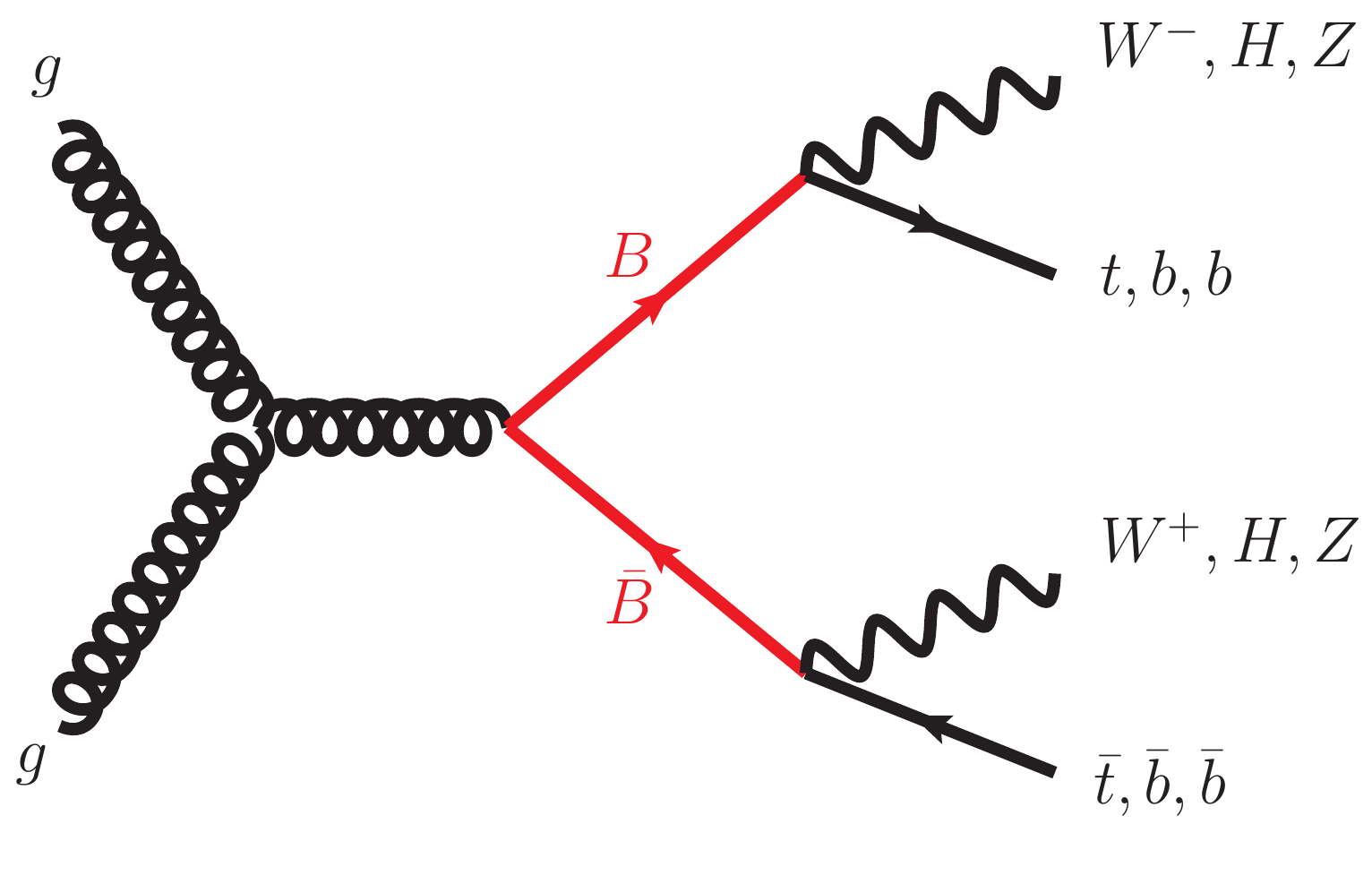}}
\caption{Representative leading-order Feynman diagrams for (a)
$T\bar{T}$ and (b) $B\bar{B}$ pair production. The studied VLQ
decays are also displayed.}
\label{fig:VLQ_FD}
\end{center}
\end{figure}

At the Large Hadron Collider (LHC), VLQs with masses below
approximately $1~\tev$ would mainly be pair-produced, a process dominated by the
strong interaction. The corresponding predicted
cross section ranges from 195~fb to 2.0~fb for quark masses from
$800~\gev$ to $1500~\gev$~\cite{Czakon:2011xx} and depends only on the
quark mass.  Production of single VLQs via the electroweak interaction
is also possible, but depends on the strength of the interaction
between the new quarks and the weak gauge bosons. Representative
Feynman diagrams for \BBbar and \TTbar
production and decay are shown in Figure~\ref{fig:VLQ_FD}.
 
The branching ratio (\BR) for each decay mode ($T$$\to$$Wb,Zt,Ht$ and
$B$$\to$$Wt,Zb,Hb$) depends on the VLQ mass and weak-isospin quantum
numbers, as calculated in Ref.~\cite{AguilarSaavedra:2009es}.  For a
singlet $T$, all three decay modes have sizable branching ratios,
while the charged-current decay mode $T$$\to$$Wb$ is absent if $T$ is
either in a $(X,T)$ doublet, where $X$ is a VLQ with a charge of
$+5/3$, or in a $(T,B)$ doublet with $|V_{Tb}| \ll |V_{tB}|$, where
$V_{ij}$ are the elements of a generalized Cabibbo--Kobayashi--Maskawa
matrix~\cite{PhysRevLett.10.531,Kobayashi:1973fv,AguilarSaavedra:2009es}.
Since the $T$ quark branching
ratios are identical in both doublets, no distinction is made between
them when referring to the doublet $T$ results.  A singlet $B$ will
have a sizable branching ratio to all three decay channels, while the
branching ratios in the doublet case depend on whether it is in a
$(T,B)$ doublet or $(B,Y)$ doublet, where $Y$ is a VLQ with a charge
of $-4/3$.  For a $(B, Y)$ doublet, only neutral current couplings to
SM quarks are allowed at leading order (LO), so the $B$$\to$$Wt$ decay is
forbidden.  Conversely, for a $(T,B)$ doublet with $|V_{Tb}| \ll
|V_{tB}|$, $B$$\to$$Wt$ is the only allowed decay.  Therefore, the
specific $B$ doublet scenario will be stated when interpreting the
results.
 
\section{Contributing analyses}
 
Searches for pair-produced VLQ partners of the third-generation quarks
have been performed by ATLAS~\cite{HTXPAP, WbXPAP, WTXPAP, ZtMETPAP,
ZTMLPAP, SSLPAP, HADPAP} and
CMS~\cite{Sirunyan:2017usq,Sirunyan:2017pks,Sirunyan:2018omb} at the
LHC at $\sqrt{s}=13~\tev$.  This Letter presents the full
combination of the ATLAS searches using 36.1~fb$^{-1}$ of data
collected in 2015 and 2016. The ATLAS detector is described in
Ref.~\cite{PERF-2007-01}. Below is a brief description of each
contributing analysis.

`\underline{\htx}'~\cite{HTXPAP}: The primary targets of this analysis
are \TTbar events with at least one VLQ decaying into $Ht$, with $H\to
b\bar{b}$.  Events must have at least six jets~\cite{Aaboud:2017jcu}
and either one lepton (electron~\cite{PERF-2013-05} or
muon~\cite{PERF-2015-10}) or missing transverse
momentum~\cite{PERF-2016-07} $\met > 200~$\GeV{} with zero leptons.
The analysis uses $b$-tagging~\cite{PERF-2012-04,
ATL-PHYS-PUB-2016-012} as well as dedicated top and Higgs jet tagging
to classify the events into 22 and 12 search regions for the
zero-lepton and one-lepton selections, respectively.  The final
discriminant is the scalar sum ($S_\text{T}$) of the transverse
momenta of the selected jets, lepton, and missing transverse
momentum. The dominant background is the associated production of
a \ttbar pair with $b$- and $c$-quark jets, which is modeled via Monte
Carlo (MC) simulation and assigned dedicated modeling uncertainties.

`\underline{\wbx}'~\cite{WbXPAP}: This analysis primarily targets
$\TTbar$$\to$$WbWb$ events with one $W$ decaying leptonically and the
other hadronically.  Event selection requires one lepton, $\geq$3
jets, at least one of them being $b$-tagged, and a
hadronically decaying $W$ boson identified using jet substructure
techniques~\cite{ATL-PHYS-PUB-2015-033}.  The final discriminant is
the reconstructed mass of the $T$$\to$$Wb$$\to$$\ell\nu b$ candidate.  The
dominant background is from $t\bar{t}$ pair production, which is
modeled using MC simulation with dedicated modeling uncertainties.
 
`\underline{\wtx}'~\cite{WTXPAP}: Very similar to the `{\wbx'} analysis,
this analysis is optimized to target \BBbar{} signals, especially in
the case where $B$$\to$$Wt$. This analysis discriminates between the signal
and the dominant \ttbar{} background in the signal regions using either a
boosted decision tree discriminant or the reconstructed mass of the $B$
candidate.
 
`\underline{\ztmet}'~\cite{ZtMETPAP}: This analysis targets
$\TTbar$$\to$$ZtZt$ events with an invisible $Z$ decay.  Events must
have $\met>300~$\GeV{}, one charged lepton from the decay of a top
quark, and $\ge 4$ small-radius jets, which are
reclustered~\cite{Nachman:2014kla} into large-radius jets.  The
analysis defines a single-bin signal region that capitalizes on
various \met{}-based variables and requires at least two high-mass
large-radius jets due to hadronically decaying top quarks and/or heavy
bosons from the VLQ decays.  The dominant backgrounds are
$t\bar{t}$+jets, $W$+jets and single-top events, which are estimated
from MC simulation and normalized using dedicated control regions.
 
`\underline{\ztll}'~\cite{ZTMLPAP}: This analysis searches for
\TTbar and \BBbar events containing a leptonically decaying $Z$
boson ($Z\to\ell^+\ell^-$) and at least two $b$-jets.  The analysis
has one trilepton signal region and three dilepton signal regions,
depending on the number of large-radius jets (0, 1, or $\ge$2).  The
final discriminant depends on the signal region.  The dominant
backgrounds for the dilepton channels are $Z$+jets and/or \ttbar{} and
diboson, while the trilepton channels are dominated by diboson ($WZ$)
and $t\bar{t}Z$ events, each modeled by MC simulation and validated
with dedicated control regions.

`\underline{Trilepton/same-sign dilepton}'~\cite{SSLPAP}: This
analysis targets \TTbar and \BBbar decays with multilepton final
states, with particular emphasis on events containing a pair of
charged leptons with the same electric charge (``same-sign'').  Eight
single-bin signal regions are defined in accord with the number of
leptons and $b$-tagged jets.  The background composition for this
analysis varies between signal regions.  Contributions from
instrumental backgrounds (fake/non-prompt leptons and electrons with
incorrectly measured charge) are estimated using data-driven
techniques, while background processes with prompt leptons,
originating mostly from $\ttbar+W$ and diboson events, are modeled
with MC simulations.
 
`\underline{Fully hadronic}'~\cite{HADPAP}: This analysis focuses on
final states with zero leptons, low \met{}, at least four
(small-radius) high-\pt{} jets, and at least two $b$-tagged jets.
This is the only analysis with significant sensitivity to $\BBbar \to
HbH\bar{b}$.  Small-radius jets are reclustered into large-radius
jets, which may be identified as top quarks, $W/Z$, or $H$ bosons
using a multi-class deep neural network~\cite{SCHMIDHUBER201585}.  The
final discriminant is the distribution of the signal likelihood
calculated using the matrix-element method~\cite{JPSJ.57.4126}.  The
dominant background is from multijet production, which is estimated
using a data-driven technique.

\begin{table}
\caption{The most sensitive decay channel for each analysis entering
the combination.  A `-' indicates that the analysis was not used
for that signal process.}
\label{tab:analyses}
\centering
\begin{tabular}{l c c}
\toprule
Analysis & $\TTbar$ decay & $\BBbar$ decay \\
\midrule
\htx~\cite{HTXPAP} & $HtH\bar{t}$ & - \\
\wbx~\cite{WbXPAP} & $WbW\bar{b}$ & - \\
\wtx~\cite{WTXPAP} & - & $WtW\bar{t}$ \\
\ztmet~\cite{ZtMETPAP} & $ZtZ\bar{t}$ & - \\
\ztll~\cite{ZTMLPAP} & $ZtZ\bar{t}$ & $ZbZ\bar{b}$ \\
Tril./s.s. dilepton~\cite{SSLPAP} & $HtH\bar{t}$ & $WtW\bar{t}$ \\
Fully hadronic~\cite{HADPAP} & $HtH\bar{t}$ & $HbH\bar{b}$ \\
\bottomrule
\end{tabular}
\end{table}
 
Most of the analyses were designed to be complementary. While each analysis
provides sensitivity to various decay configurations, the most sensitive
is shown in Table~\ref{tab:analyses}. All analyses use
consistent definitions for the reconstructed physics objects, so only
a few additional selection requirements were needed to suppress
overlap.  Compared to the standalone analyses, the \wbx{} and \ztmet{}
analyses removed events with $\ge$6 jets and $\ge$3 $b$-jets to
avoid overlap with the \htx{} selection.  The \ztmet{} analysis also
requires $S_\text{T} < 1.8$~\TeV{} in a control region to mitigate the
overlap with a signal region in the \wbx{} analysis.  To reduce overlap
with the
\ztll analysis, the trilepton/same-sign dilepton analysis removed
events with more than three leptons or events with a lepton pair
having an invariant mass compatible with a $Z$ boson ($Z$-veto).  This
$Z$-veto is the only added selection requirement with significant
impact on the individual analysis sensitivity; however, that
sensitivity is recovered by the \ztll analysis.  After applying these
additional selection requirements, the fraction of events falling into
more than one analysis region was evaluated to be less than 1\%
between any two signal regions and less than 3\% between any pair of
signal or control regions and has negligible impact on the results.

The VLQ signal samples used by the analyses were generated with the LO
generator {\textsc{Protos}}~v2.2~\cite{protos} using the NNPDF2.3
LO~\cite{Ball:2012cx} set of parton distribution functions (PDF) and
passed to {\textsc{Pythia}} 8.186~\cite{Sjostrand:2007gs} for parton
showering and fragmentation.  The samples are normalized using
cross-sections computed with {\textsc{ Top++}}
v2.0~\cite{Czakon:2011xx} at next-to-next-to-leading order (NNLO) in
QCD, including resummation of next-to-next-to-leading logarithmic soft
gluon
terms~\cite{Cacciari:2011hy,Baernreuther:2012ws,Czakon:2012zr,Czakon:2012pz,Czakon:2013goa},
and using the MSTW 2008 NNLO~\cite{Martin:2009iq,Martin:2009bu}
PDF. Further information about simulated events and details of the background
estimations for each analysis can be found in the respective
publications.

\section{Statistical analysis}
 
The statistical analysis is the same as in the individual analyses and
is based on a binned likelihood function 
constructed as the product of the Poisson probabilities of all bins
entering the combination.  This function depends on the
signal-strength parameter $\mu$, a factor multiplying the
theoretical signal cross-section ($\mu\equiv\sigma/\sigma_\text{theory}$),
and 
a set of nuisance parameters that encode the effect of the
systematic uncertainties on the signal and background expectations. These
parameters are included with Gaussian or log-normal constraints.
Additional unconstrained nuisance parameters are included to control the
normalization of the main backgrounds, following the settings used in
the standalone searches. The combination is achieved by performing a
fit with all bins from all the regions considered from each
analysis.
 
The analysis is limited by statistical uncertainties, and the precise
correlation model for the systematic uncertainties was found to not
significantly affect the results.  The detector-related uncertainties
are treated as fully correlated across analyses, with the following
exceptions.  The central values and uncertainties of the $b$-tagging
and the luminosity measurement were updated after the publication of
the $\ztmet$ and $\wbx$ analyses.  Therefore, to avoid propagating
constraints caused by the change in the method, these uncertainties
are correlated between the $\ztmet$ and $\wbx$ analyses, but
uncorrelated with the other searches, which are correlated among
themselves.
The modeling uncertainties and background normalization parameters are treated
as uncorrelated between analyses. Although some background processes are
common to multiple analyses, the phase space and the techniques used to
estimate those backgrounds can be quite different.  Residual correlations
are therefore expected to be negligible.

\section{Results}

The behavior of the combination is consistent with the fits from the
individual analyses.  The post-fit values of all nuisance parameters
are compatible with the standalone analyses, with the constraints
generally determined by the analysis most sensitive to the given
nuisance parameter.  Similarly, the background predictions in each
analysis after the combined fit are very close to the results from the
standalone analyses.  After the combination, no significant excess is
observed in the data, so 95\% confidence level (CL) limits are set on
the cross-section of a VLQ
signal.  To increase the applicability and usefulness of this
combination, limits are evaluated both for benchmark scenarios with
specific branching ratios and for general combinations of branching
ratios.

\begin{figure}[htb]
\centering
\subfigure[]{\includegraphics[width=0.45\textwidth]{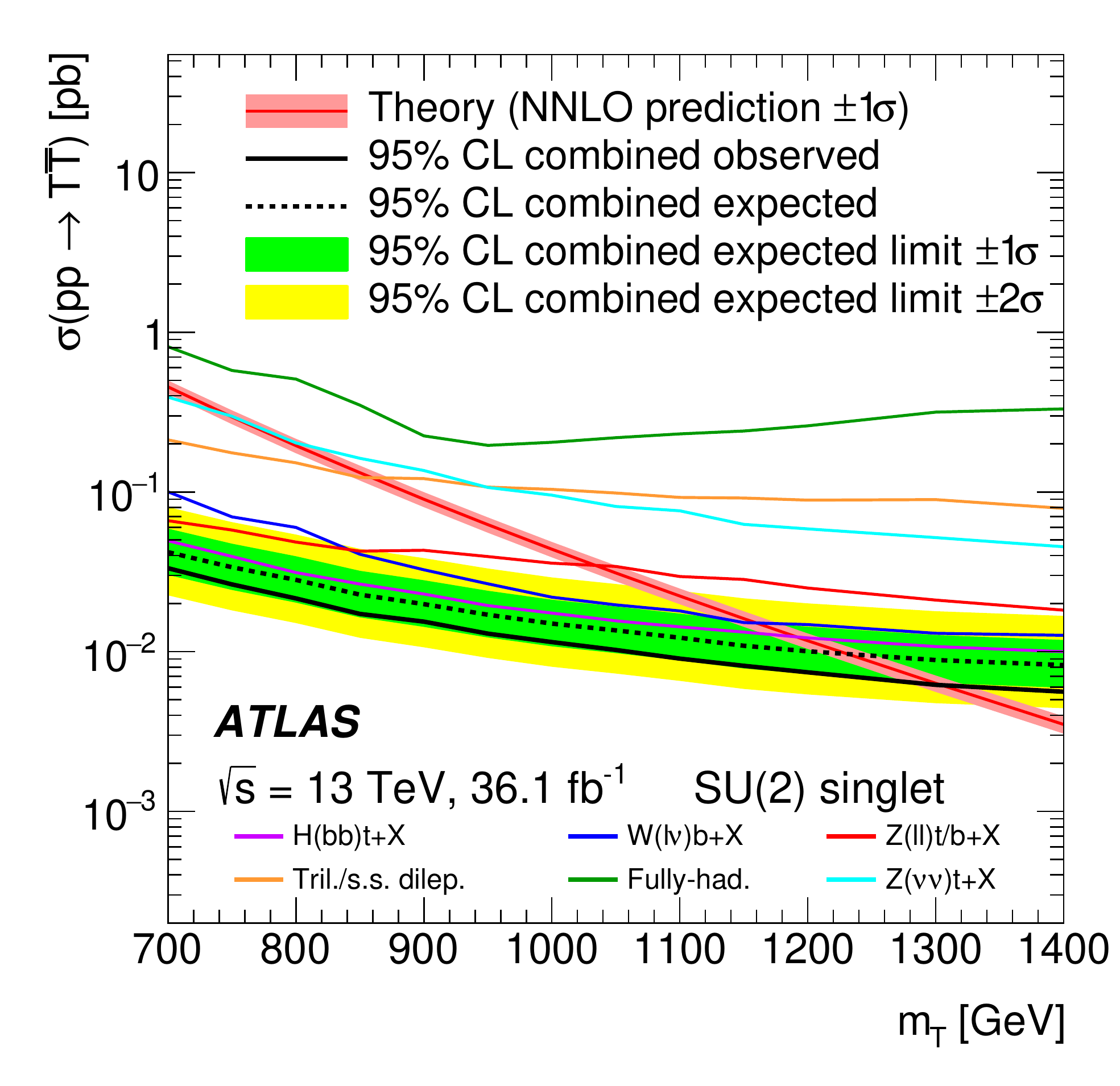}}
\subfigure[]{\includegraphics[width=0.45\textwidth]{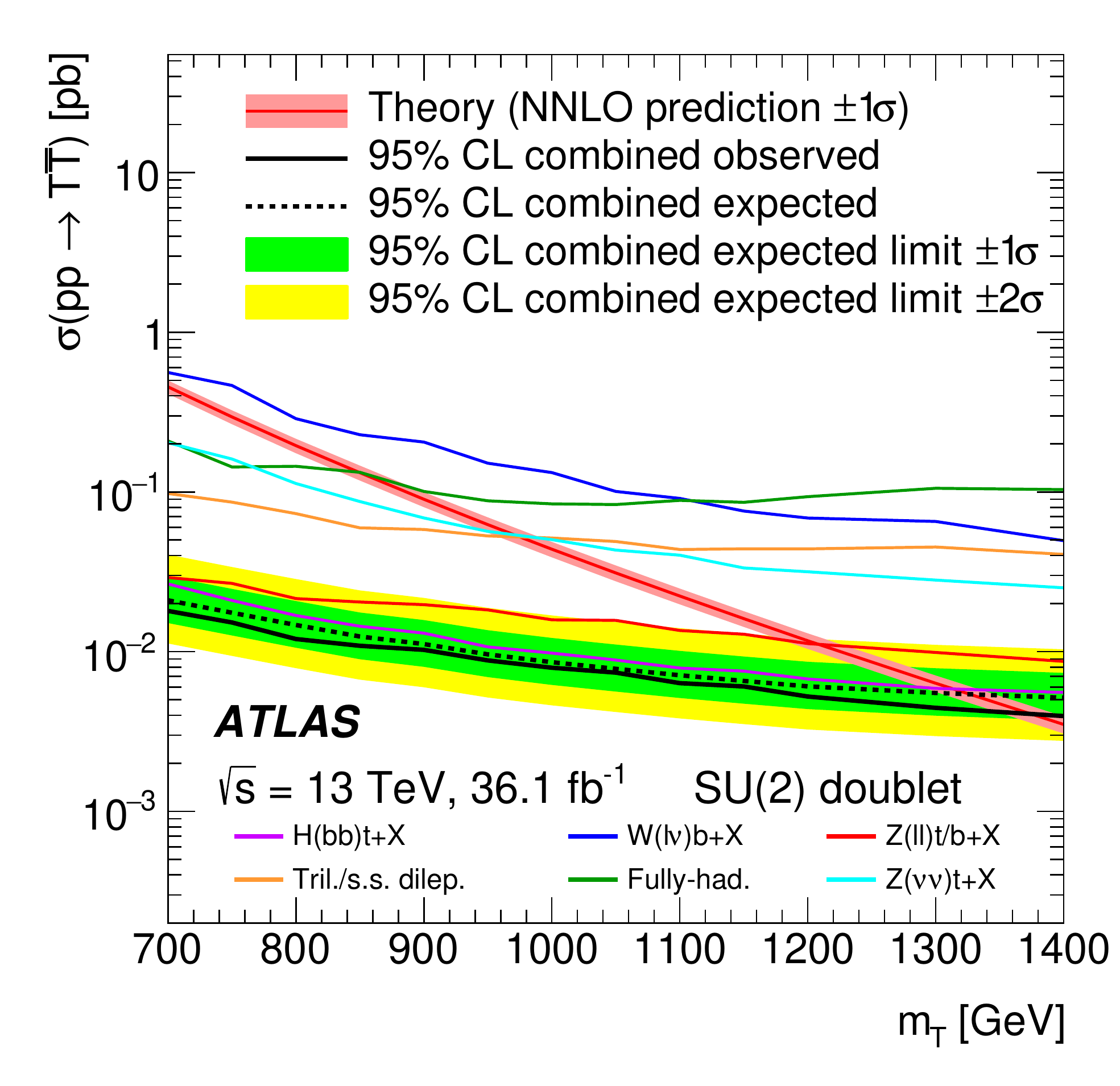}}
\caption{Observed (solid lines) and expected (dashed line) 95\% CL
upper limits on the \TTbar cross-section versus mass for the
combination and the standalone analyses in black and colored lines,
respectively. The (a) singlet and (b) doublet scenarios~\cite{AguilarSaavedra:2009es} are displayed. The shaded bands correspond to
$\pm1$ and $\pm2$ standard deviations around the combined expected
limit.  The rapidly falling thin red line and band show the theory prediction and
corresponding uncertainty~\cite{Czakon:2011xx}, respectively.  }
\label{fig:limits_1D_T}
\end{figure}

\begin{figure}[htb]
\centering
\subfigure[]{\includegraphics[width=0.45\textwidth]{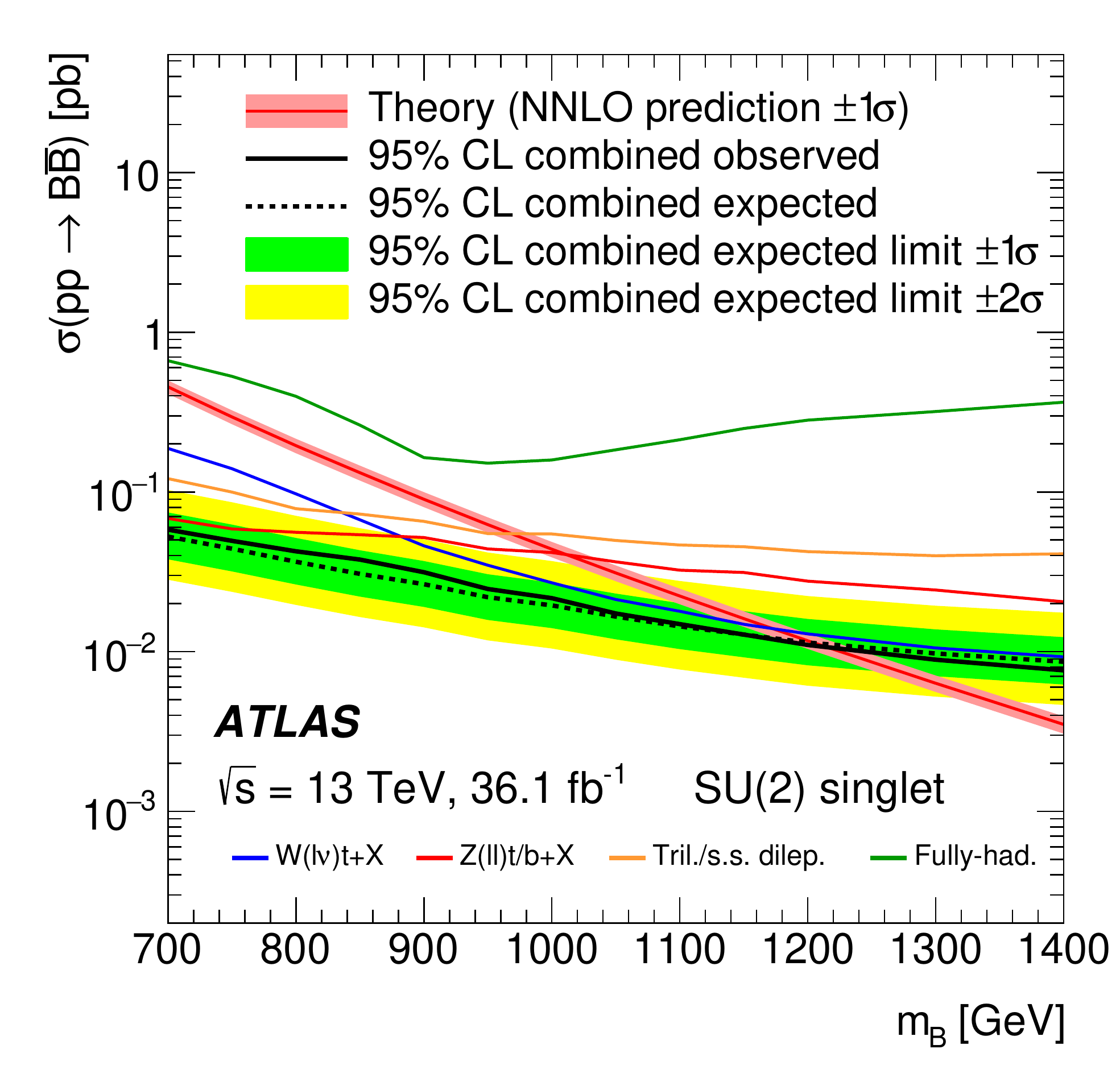}}
\subfigure[]{\includegraphics[width=0.45\textwidth]{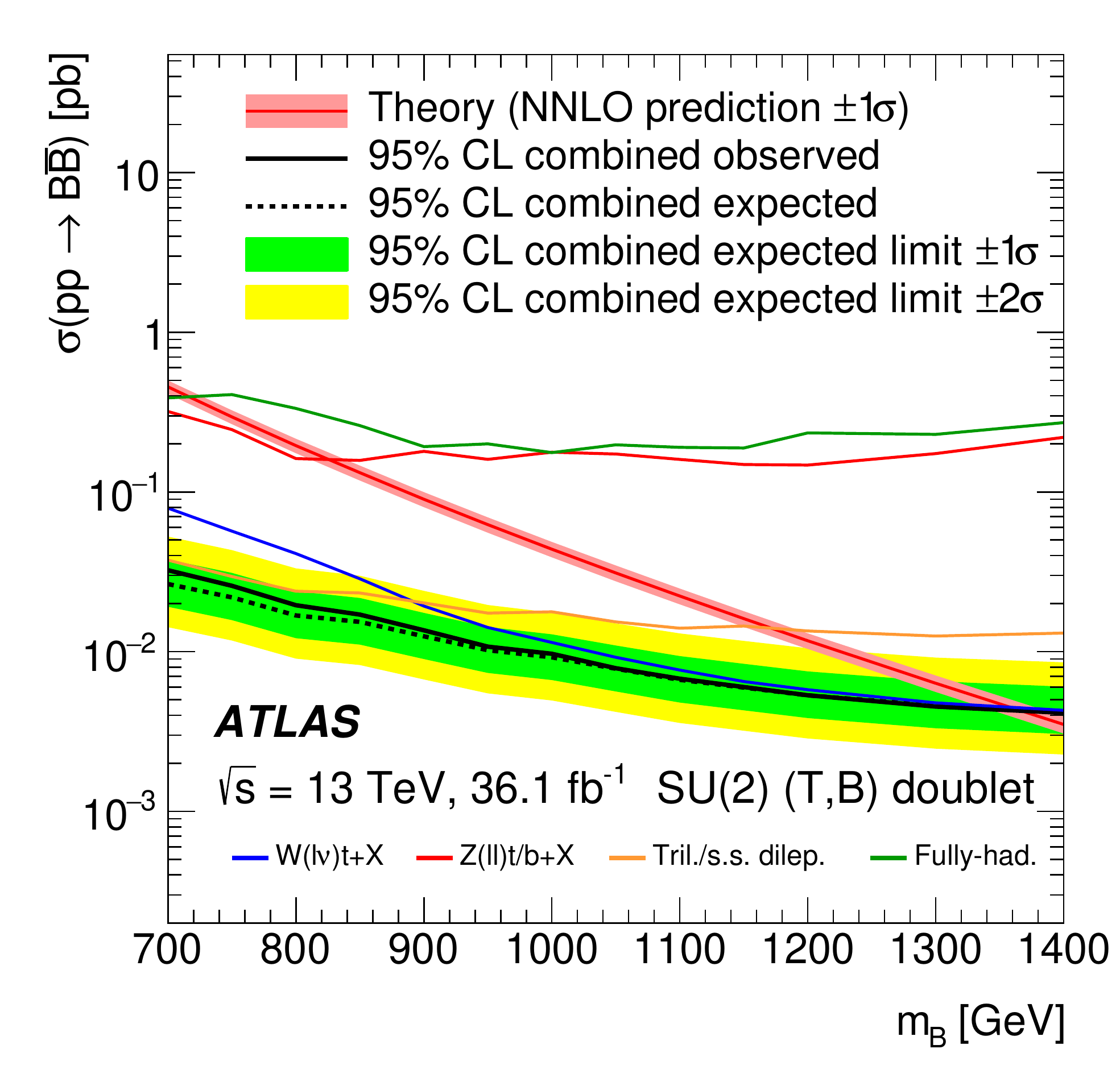}}
\subfigure[]{\includegraphics[width=0.49\textwidth]{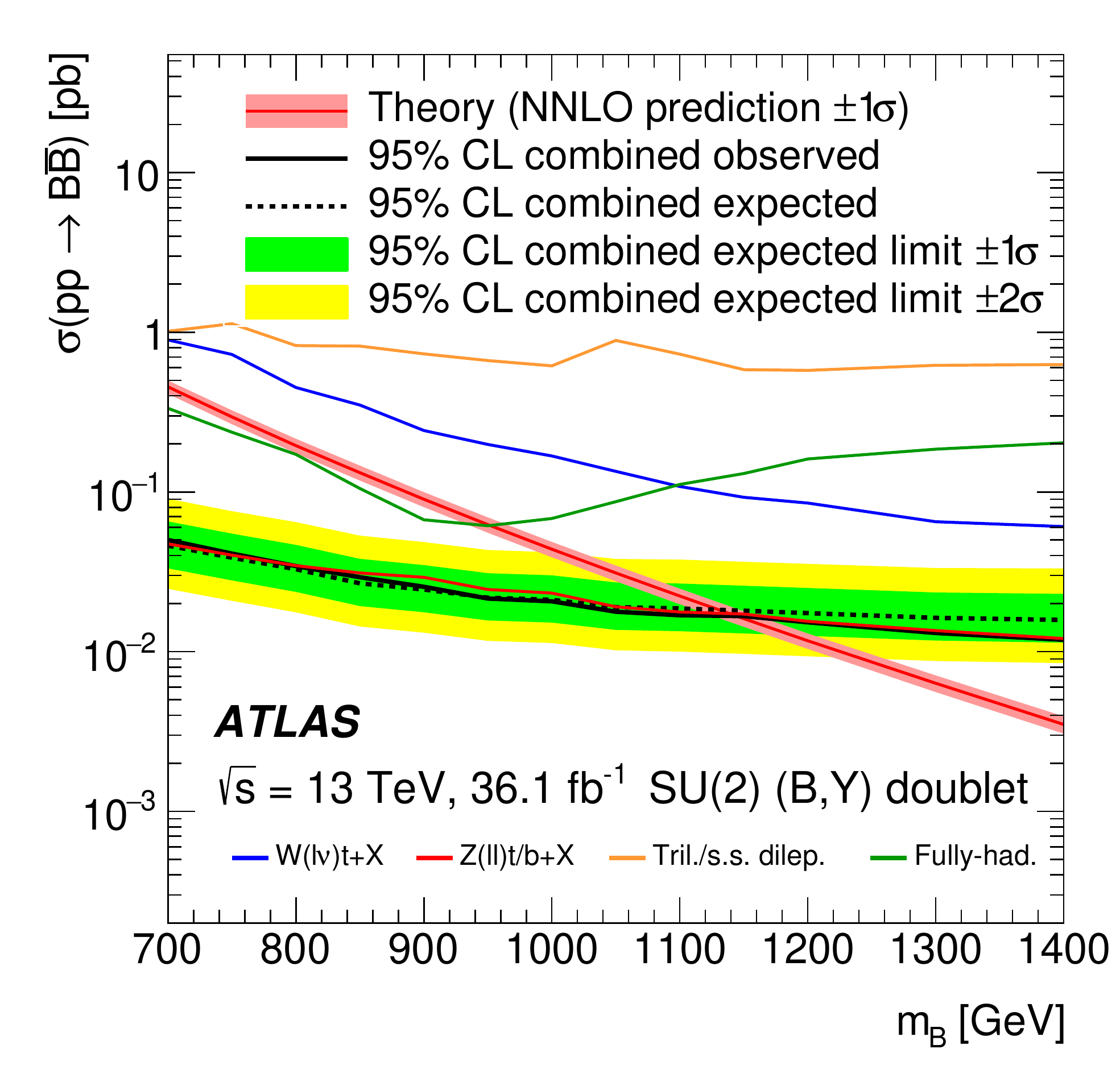}}
\caption{Observed (solid lines) and expected (dashed line) 95\% CL
upper limits on the \BBbar cross-section versus mass for the
combination and the standalone analyses in black and colored lines,
respectively. The (a) singlet, (b) $(T,B)$ doublet, and (c) $(B,Y)$ doublet
scenarios~\cite{AguilarSaavedra:2009es} are displayed. The shaded bands
correspond to $\pm1$ and $\pm2$ standard deviations around the
combined expected limit.  The rapidly falling thin red line and band show the
theory prediction and corresponding uncertainty~\cite{Czakon:2011xx}, respectively.  }
\label{fig:limits_1D_B}
\end{figure}

For an assumed set of branching ratios, upper limits are set on the
production cross-sections for $\TTbar$ and $\BBbar$ as a function of
the VLQ mass using the \cls{} method~\cite{Junk:1999kv,Read:2002hq}
with the asymptotic approximation~\cite{Cowan:2010js}.  Observed and
expected upper limits on the \TTbar cross-sections as a function of
mass are shown in Figure~\ref{fig:limits_1D_T} for the benchmark
scenarios of an isospin singlet or doublet $T$.  Analogous limits on
the \BBbar cross-section are shown in Figure~\ref{fig:limits_1D_B}.
The observed
limits from the individual analyses, after the additional selections
defined in this letter, are also shown.
For a singlet $T$, masses below \TLimitSinglet{}\,\tev{} are excluded,
while a $T$ in an isospin doublet is excluded for masses below
\TLimitDoublet{}\,\tev{}.  A singlet $B$ is excluded for masses below
\BLimitSinglet{}\,\tev, a $B$ in a $(T,B)$ doublet is excluded for masses
below \BLimitTDoublet{}\,\tev{}, and a $B$ in a $(B,Y)$ doublet is
excluded for masses below \BLimitYDoublet{}\,\tev{}.
 
The combination is
significantly more sensitive than any one analysis.  For example, in
the case of the SU(2) singlet, the observed limit on the \TTbar cross-section
is improved by up to a factor of $\sim 1.7$, which translates to an
increase of 110~\GeV{} in the observed mass limit.

\begin{figure}[htb]
\centering
\subfigure[]{\includegraphics[width=0.49\textwidth]{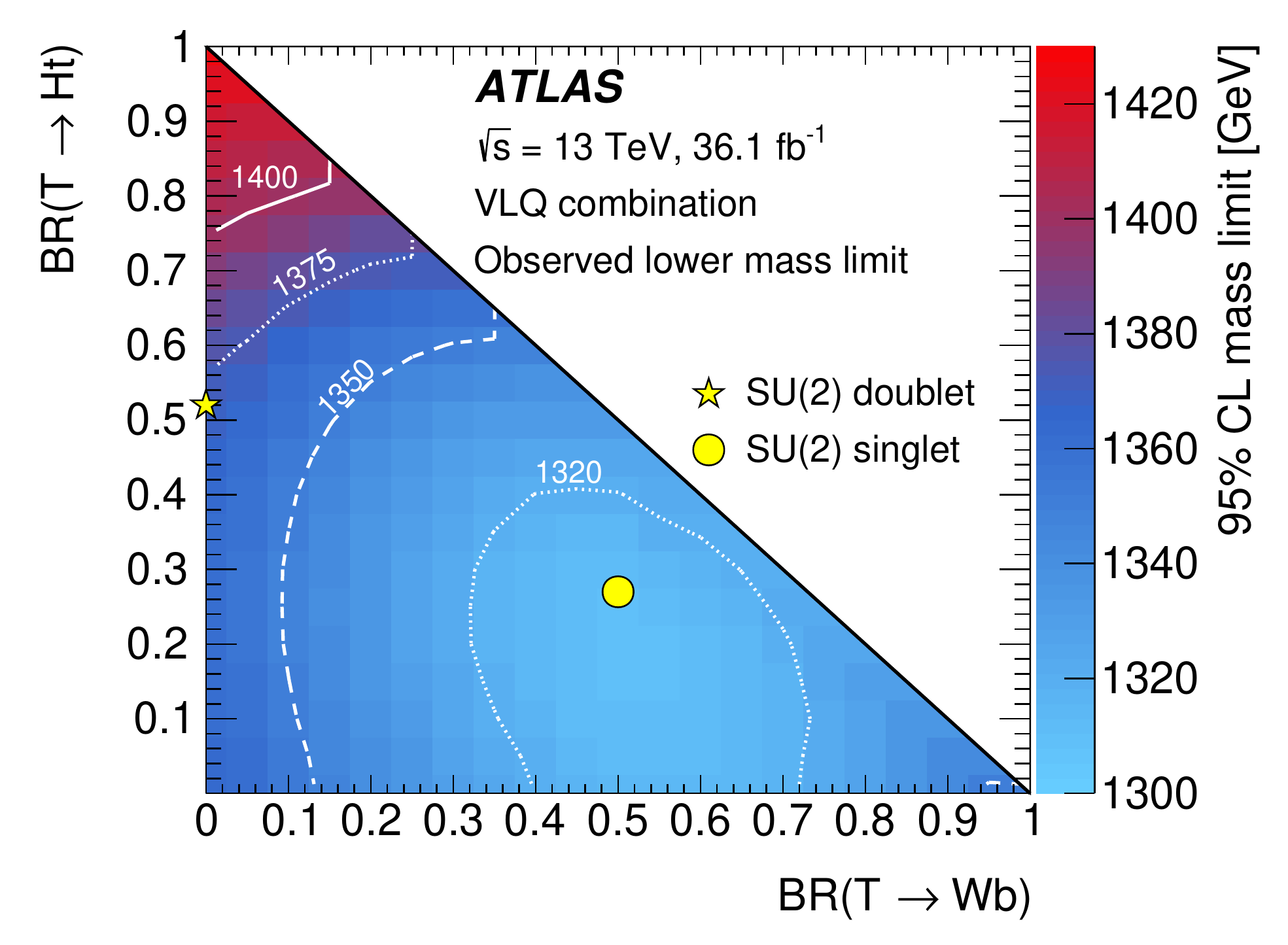}}
\subfigure[]{\includegraphics[width=0.49\textwidth]{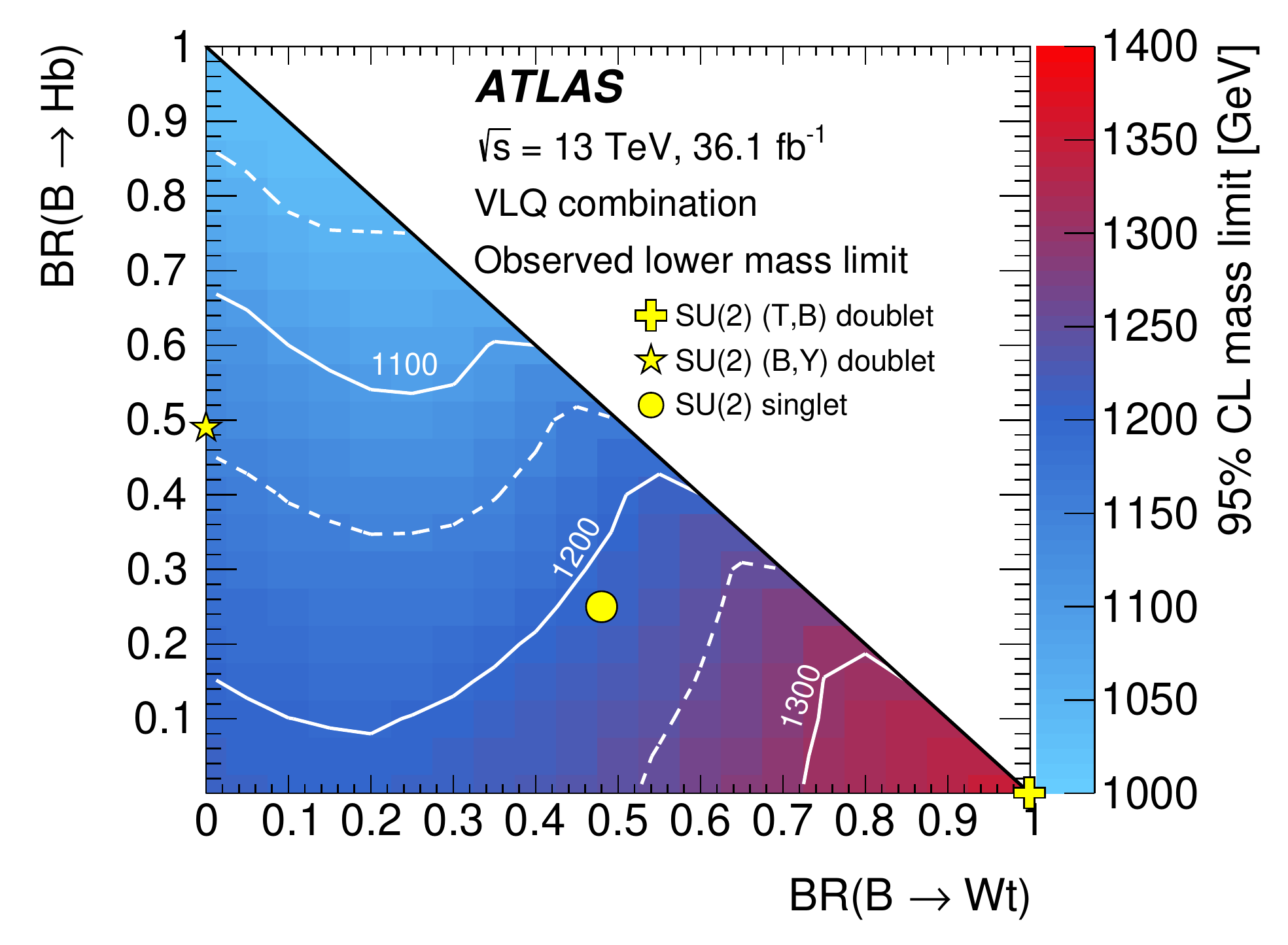}}
\caption{ Observed lower limits at 95\% CL on the mass of the (a)
$T$ and (b) $B$ as a function of branching ratio assuming
$\text{\BR}(T$$\to$$Ht) + \text{\BR}(T$$\to$$Zt) + \text{\BR}(T \to
Wb) = 1$ and $\text{\BR}(B$$\to$$Hb) + \text{\BR}(B$$\to$$Zb) +
\text{\BR}(B$$\to$$Wt) = 1$. The yellow markers
indicate the branching ratios for the SU(2) singlet and doublet
scenarios
where the branching ratios become
approximately independent of the VLQ mass~\cite{AguilarSaavedra:2009es}.
} \label{fig:limits_2D_temp}
\end{figure}

In addition, model-independent lower limits are set on the VLQ mass
for all combinations of branching ratios, assuming
$\text{\BR}(T$$\to$$Ht) + \text{\BR}(T$$\to$$Zt) +
\text{\BR}(T$$\to$$Wb) = 1$ and $\text{\BR}(B$$\to$$Hb) +
\text{\BR}(B$$\to$$Zb) + \text{\BR}(B$$\to$$Wt) = 1$.  The resulting
lower limits on the VLQ mass as a function of branching ratio are
presented in Figure~\ref{fig:limits_2D_temp}.  Limits corresponding to
$\text{\BR}(T$$\to$$Wb)=1$ and $\text{\BR}(B$$\to$$Wt)=1$ are found to
also be applicable to $Y\bar{Y}$$\to$$WbWb$ and $X\bar{X}$$\to$$WtWt$,
respectively.
The high degree of complementarity between the analyses is clearly
demonstrated in Figure~\ref{fig:limits_2D_temp}.  For {\em any}
combination of branching ratios, the combined analysis leads to observed
(expected) lower mass limits of \TLimitMin~(\TExpLimitMin)~\tev{} for
$T$ and \BLimitMin~(\BExpLimitMin)~\tev{} for $B$.
Limits on the signal strength, which can be
used to interpret the results in scenarios with additional VLQ decays
that escape detection~\cite{Ballesteros:2017xeg}, are available in the
HEPData repository~\cite{hepdata,comboHEPData}.

\FloatBarrier
 
\section{Conclusion}
 
The ATLAS Collaboration has performed a combination of seven analyses searching
for
pair-produced VLQs.  Upper limits on the
cross-section are determined and used to set lower limits on the VLQ
mass for various benchmark scenarios and for general combinations of
branching ratios.  This combination results in the most stringent
limits to date on VLQ pair production.  Due to the high degree of
complementarity between the analyses, the combination has
significantly better sensitivity than the standalone
analyses, for the first time excluding $T$ ($B$) masses below
\TLimitMin{} (\BLimitMin)~\tev{} for {\em any} combination of decays
into SM particles.

\section*{Acknowledgments}
 
% The next lines are included from the .//acknowledgements/Acknowledgements.tex input file
 
We thank CERN for the very successful operation of the LHC, as well as the
support staff from our institutions without whom ATLAS could not be
operated efficiently.
 
We acknowledge the support of ANPCyT, Argentina; YerPhI, Armenia; ARC, Australia; BMWFW and FWF, Austria; ANAS, Azerbaijan; SSTC, Belarus; CNPq and FAPESP, Brazil; NSERC, NRC and CFI, Canada; CERN; CONICYT, Chile; CAS, MOST and NSFC, China; COLCIENCIAS, Colombia; MSMT CR, MPO CR and VSC CR, Czech Republic; DNRF and DNSRC, Denmark; IN2P3-CNRS, CEA-DRF/IRFU, France; SRNSFG, Georgia; BMBF, HGF, and MPG, Germany; GSRT, Greece; RGC, Hong Kong SAR, China; ISF and Benoziyo Center, Israel; INFN, Italy; MEXT and JSPS, Japan; CNRST, Morocco; NWO, Netherlands; RCN, Norway; MNiSW and NCN, Poland; FCT, Portugal; MNE/IFA, Romania; MES of Russia and NRC KI, Russian Federation; JINR; MESTD, Serbia; MSSR, Slovakia; ARRS and MIZ\v{S}, Slovenia; DST/NRF, South Africa; MINECO, Spain; SRC and Wallenberg Foundation, Sweden; SERI, SNSF and Cantons of Bern and Geneva, Switzerland; MOST, Taiwan; TAEK, Turkey; STFC, United Kingdom; DOE and NSF, United States of America. In addition, individual groups and members have received support from BCKDF, CANARIE, CRC and Compute Canada, Canada; COST, ERC, ERDF, Horizon 2020, and Marie Sk{\l}odowska-Curie Actions, European Union; Investissements d' Avenir Labex and Idex, ANR, France; DFG and AvH Foundation, Germany; Herakleitos, Thales and Aristeia programmes co-financed by EU-ESF and the Greek NSRF, Greece; BSF-NSF and GIF, Israel; CERCA Programme Generalitat de Catalunya, Spain; The Royal Society and Leverhulme Trust, United Kingdom.
 
The crucial computing support from all WLCG partners is acknowledged gratefully, in particular from CERN, the ATLAS Tier-1 facilities at TRIUMF (Canada), NDGF (Denmark, Norway, Sweden), CC-IN2P3 (France), KIT/GridKA (Germany), INFN-CNAF (Italy), NL-T1 (Netherlands), PIC (Spain), ASGC (Taiwan), RAL (UK) and BNL (USA), the Tier-2 facilities worldwide and large non-WLCG resource providers. Major contributors of computing resources are listed in Ref.~\cite{ATL-GEN-PUB-2016-002}.
 
% End of text imported from the .//acknowledgements/Acknowledgements.tex input file

\clearpage
 
\printbibliography
 
\onecolumn
 
\include{atlas_authlist} 
\end{document}

%% file: atlas_authlist.tex
% ATLAS Collaboration author list
% Reference date of EXOT-2017-17 is 2018-06-16
% Author list last updated on date 24-JUL-18
% Data extracted on 24-Jul-2018 for paper reference EXOT-2017-17
% at 4:41pm
 
\begin{flushleft}
{\Large The ATLAS Collaboration}

\bigskip

M.~Aaboud$^\textrm{\scriptsize 34d}$,    
G.~Aad$^\textrm{\scriptsize 99}$,    
B.~Abbott$^\textrm{\scriptsize 124}$,    
O.~Abdinov$^\textrm{\scriptsize 13,*}$,    
B.~Abeloos$^\textrm{\scriptsize 128}$,    
D.K.~Abhayasinghe$^\textrm{\scriptsize 91}$,    
S.H.~Abidi$^\textrm{\scriptsize 164}$,    
O.S.~AbouZeid$^\textrm{\scriptsize 39}$,    
N.L.~Abraham$^\textrm{\scriptsize 153}$,    
H.~Abramowicz$^\textrm{\scriptsize 158}$
% ...
\bigskip
\\

% ...
$^{13}$Institute of Physics, Azerbaijan Academy of Sciences, Baku; Azerbaijan.\\
% ...
$^{34}$$^{(a)}$Facult\'e des Sciences Ain Chock, R\'eseau Universitaire de Physique des Hautes Energies - Universit\'e Hassan II, Casablanca;$^{(b)}$Centre National de l'Energie des Sciences Techniques Nucleaires (CNESTEN), Rabat;$^{(c)}$Facult\'e des Sciences Semlalia, Universit\'e Cadi Ayyad, LPHEA-Marrakech;$^{(d)}$Facult\'e des Sciences, Universit\'e Mohamed Premier and LPTPM, Oujda;$^{(e)}$Facult\'e des sciences, Universit\'e Mohammed V, Rabat; Morocco.\\
% ...
$^{39}$Niels Bohr Institute, University of Copenhagen, Copenhagen; Denmark.\\
% ...
$^{91}$Department of Physics, Royal Holloway University of London, Egham; United Kingdom.\\
% ...
$^{99}$CPPM, Aix-Marseille Universit\'e, CNRS/IN2P3, Marseille; France.\\
% ...
$^{124}$Homer L. Dodge Department of Physics and Astronomy, University of Oklahoma, Norman OK; United States of America.\\
% ...
$^{128}$LAL, Universit\'e Paris-Sud, CNRS/IN2P3, Universit\'e Paris-Saclay, Orsay; France.\\
% ...
$^{153}$Department of Physics and Astronomy, University of Sussex, Brighton; United Kingdom.\\
% ...
$^{158}$Raymond and Beverly Sackler School of Physics and Astronomy, Tel Aviv University, Tel Aviv; Israel.\\
% ...
$^{164}$Department of Physics, University of Toronto, Toronto ON; Canada.\\
% ...
$^{a}$ Also at Borough of Manhattan Community College, City University of New York, NY; United States of America.\\
$^{b}$ Also at Centre for High Performance Computing, CSIR Campus, Rosebank, Cape Town; South Africa.\\
$^{c}$ Also at CERN, Geneva; Switzerland.\\
$^{d}$ Also at CPPM, Aix-Marseille Universit\'e, CNRS/IN2P3, Marseille; France.\\
$^{e}$ Also at D\'epartement de Physique Nucl\'eaire et Corpusculaire, Universit\'e de Gen\`eve, Gen\`eve; Switzerland.\\
% ...
$^{*}$ Deceased

\end{flushleft}

% Created with Glance <Atlas.Glance@cern.ch>